\documentclass[pra,aps,nofootinbib,notitlepage,twocolumn]{revtex4-1}

\usepackage{amsthm}
\usepackage{amsmath}
\usepackage{amssymb}
\usepackage{amsfonts}
\usepackage{graphicx}

\usepackage{txfonts}

\usepackage{ulem}
\usepackage{verbatim}

\usepackage[colorlinks=true,linkcolor=blue,citecolor=blue,urlcolor=blue]{hyperref}

\begin{document}

\title{Enhancement of the metrological sensitivity limit through knowledge of the average energy}
\author{Manuel Gessner}
\email{manuel.gessner@ens.fr}
\affiliation{D\'{e}partement de Physique, \'{E}cole Normale Sup\'{e}rieure, PSL Universit\'{e}, CNRS,
24 Rue Lhomond, 75005 Paris, France}
\affiliation{Laboratoire Kastler Brossel, ENS-PSL, CNRS, Sorbonne Universit\'{e}, Coll\`{e}ge de France, 24 Rue Lhomond, 75005 Paris, France}
\date{\today}

\begin{abstract}
We consider the problem of quantum phase estimation with access to arbitrary measurements in a single suboptimal basis. The achievable sensitivity limit in this case is determined by the classical Cram\'{e}r-Rao bound with respect to the fixed basis. Here we show that the sensitivity can be enhanced beyond this limit if knowledge about the energy expectation value is available. The combined information is shown to be equivalent to a direct measurement of an optimal linear combination of the basis projectors and the phase-imprinting Hamiltonian. Application to an atomic clock with oversqueezed spin states yields a sensitivity gain that scales linearly with the number of atoms. Our analysis further reveals that small modifications of the observable can have a strong impact on the sensitivity.
\end{abstract}

\maketitle

\section{Introduction}
Quantum metrology aims to enhance the sensitivity of measurements by making efficient use of the properties of quantum mechanical states and measurements \cite{Helstrom,Paris,GiovannettiNATPHOT2011,Varenna,RMP}. Most of the theoretical efforts so far have focused on the identification and generation of highly sensitive quantum states \cite{RMP}, implicitly assuming that an optimal measurement can be realized. The usage of such states in metrology experiments is often challenged by their fragility under unavoidable noise processes \cite{HuelgaPRL1997,Davidovich2011,RafalNATCOMMUN2012,Acin,SmirnePRL2016}. However, even in situations where the initial state cannot be controlled, the identification of more sensitive measurement observables can be a beneficial strategy towards an improvement of the measurement precision \cite{BraunsteinPRL1994,GiovannettiPRL2006,Pezze2017,GessnerPRL2019}. For example, optimizing the measurement observable for an estimation of the separation of two incoherent light sources can overcome classical resolution limits, without the need for nonclassical sources \cite{TsangPRX2016}. Energy measurements in a fixed basis can further improve the precision for an estimation of Hamiltonian parameters \cite{SevesoPRA2018}.

Besides precision measurements, metrological sensitivity can be used as an entanglement witness by comparing to suitable sensitivity bounds for different classes of separable states \cite{PezzePRL2009,Varenna,HyllusPRA2012,GessnerPRA2016,GessnerPRA2017,GessnerPRL2018,GuehneToth,Appelaniz}. The sensitivity thus provides information about the number of entangled atoms \cite{HyllusPRA2012} and the microscopic entanglement structure in the case of individually addressable subsystems in multi-mode systems \cite{GessnerPRA2016,GessnerQuantum2017}. Metrological entanglement witnesses have been implemented successfully with Gaussian and non-Gaussian states of cold atoms \cite{RMP} and multi-mode squeezed states of light \cite{QinNPJQI2019}.

For the estimation of the phase parameter $\theta$ from any quantum state $\hat{\rho}(\theta)=e^{-i\hat{H}\theta}\hat{\rho}e^{i\hat{H}\theta}$, the optimal measurement is theoretically known and can yield a sensitivity as large as the quantum Fisher information $F_Q[\hat{\rho},\hat{H}]$ \cite{BraunsteinPRL1994}. Since the implementation of the optimal measurement may not always be feasible, it is important to identify measurement strategies that maximize the sensitivity under experimentally motivated constraints \cite{GessnerPRL2019}. Generally, to reach high sensitivities, we need observables with small variance and strong dependence on the parameter. For instance, if measurements are limited to a single, suboptimal basis, specified by the complete set of projectors $\hat{\boldsymbol{\Pi}}=(\hat{\Pi}_1,\hat{\Pi}_2,\dots,\hat{\Pi}_r)$, the maximal achievable sensitivity is given by the classical Fisher information $F[\hat{\rho}(\theta),\hat{\boldsymbol{\Pi}}]$.

In this article, we show that the achievable sensitivity is further enhanced if in addition to arbitrary measurements in the basis $\hat{\boldsymbol{\Pi}}$, the energy expectation value $\langle\hat{H}\rangle_{\hat{\rho}}$ is available, e.g., from knowledge of the initial state and the interferometer device or by calibration measurements. The enhancement occurs in spite of the fact that $\langle\hat{H}\rangle_{\hat{\rho}}$ does not depend on the parameter $\theta$ at all and is by itself an unsuitable observable for phase estimation. The scheme is equivalent to a direct measurement of an optimal observable that can be expressed as a linear combination of the elements of $\hat{\boldsymbol{\Pi}}$ and $\hat{H}$. We provide an analytical expression for the sensitivity gain that is obtained from the contribution of $\hat{H}$ to this optimal measurement observable. By applying this technique to the example of an atomic clock, we find that the sensitivity can be enhanced by a factor proportional to the number of atoms $N$. Surprisingly, this enhancement is achieved by a seemingly (but not actually) negligible contribution of $\hat{H}$ to the optimal observable. Our results further illustrate that tiny changes of the measurement observable can have dramatic effects on the sensitivity.

\section{Sensitivity gain from Hamiltonian measurements}
The method of moments, a widely used protocol for phase estimation, uses only the average value of some observable $\hat{X}$ to estimate the true value of $\theta$ \cite{RMP}. After many repeated measurements, $\mu\gg 1$, it yields an estimator variance of $(\Delta\theta_{\mathrm{est}})^2= \chi^2[\hat{\rho}(\theta),\hat{H},\hat{X}]/\mu$, where
\begin{align}\label{eq:chi}
\chi^2[\hat{\rho}(\theta),\hat{H},\hat{X}]=\frac{(\Delta\hat{X})^2_{\hat{\rho}(\theta)}}{\vert \langle [\hat{X},\hat{H}] \rangle_{\hat{\rho}(\theta)} \vert^2}.
\end{align}
It was recently shown how the observable $\hat{X}$ can be chosen in an optimal way out of some family of accessible operators \cite{GessnerPRL2019}. Consider, for example, the case of an experimental setup that provides access to measurements in one particular basis. Assuming that arbitrary observables that are diagonal in that basis, i.e., $\hat{X}=\sum_{x=1}^rc_x\hat{\Pi}_x$, can be measured, the maximal sensitivity is given by \cite{GessnerPRL2019}
\begin{align}\label{eq:ChiclassicalF}
\chi^{-2}_{\rm max}[\hat{\rho}(\theta),\hat{H},\hat{\boldsymbol{\Pi}}]:\!&=
\max_{\hat{X}\in\mathrm{span}(\hat{\boldsymbol{\Pi}})}\chi^{-2}[\hat{\rho}(\theta),\hat{H},\hat{X}]\notag\\&=F[\hat{\rho}(\theta),\hat{\boldsymbol{\Pi}}],
\end{align}
where $F[\hat{\rho}(\theta),\hat{\boldsymbol{\Pi}}]=\sum_{x=1}^rp(x|\theta)[\frac{\partial}{\partial\theta} \log p(x|\theta)]^2$ is the Fisher information with $p(x|\theta)=\mathrm{Tr}\{\hat{\rho}(\theta)\hat{\Pi}_x\}$. This maximal sensitivity is achieved by measurements of an optimally chosen linear combination of the $\hat{\boldsymbol{\Pi}}$. Equivalently, the same sensitivity can be achieved asymptotically by a maximum-likelihood estimation if the full counting statistics of individual measurement results in the basis $\hat{\boldsymbol{\Pi}}$ is available \cite{Varenna}. These measurement strategies thus saturate the Cram\'{e}r-Rao bound, which expresses that any phase estimation protocol with measurements in the basis $\hat{\boldsymbol{\Pi}}$ is limited to estimator variances of $(\Delta\theta_{\rm est})^2\geq (\Delta\theta_{\rm CR, \hat{\boldsymbol{\Pi}}})^2=\{\mu F[\hat{\rho}(\theta),\hat{\boldsymbol{\Pi}}]\}^{-1}$~\cite{Helstrom}.

By adding the generating Hamiltonian $\hat{H}$ to the set $\hat{\boldsymbol{\Pi}}$ of accessible operators, the sensitivity is further enhanced. Specifically, for $\hat{\mathbf{H}}=(\hat{H},\hat{\Pi}_1,\dots,\hat{\Pi}_r)$ we obtain the maximal sensitivity
\begin{align}\label{eq:chiQPiH}
\chi^{-2}_{\rm max}[\hat{\rho}(\theta),\hat{H},\hat{\mathbf{H}}]&=
\max_{\hat{X}\in\mathrm{span}(\hat{\mathbf{H}})}\chi^{-2}[\hat{\rho}(\theta),\hat{H},\hat{X}]\notag\\&=F[\hat{\rho}(\theta),\hat{\boldsymbol{\Pi}}]+E[\hat{\rho}(\theta),\hat{H},\hat{\boldsymbol{\Pi}}],
\end{align}
which is the central result of this article. The sensitivity enhancement $E[\hat{\rho}(\theta),\hat{H},\hat{\boldsymbol{\Pi}}]=ab^2$, with
\begin{align}
a=\left[(\Delta\hat{H})^2_{\hat{\rho}(\theta)}-\sum_{x=1}^r\frac{1}{p(x|\theta)}\mathrm{Cov}(\hat{H},\hat{\Pi}_x)^2_{\hat{\rho}(\theta)}\right]^{-1}
\end{align}
and
\begin{align}
b=\sum_{x=1}^r\mathrm{Cov}(\hat{H},\hat{\Pi}_x)_{\hat{\rho}(\theta)}\left(\frac{\partial}{\partial \theta} \log p(x|\theta)\right),
\end{align}
is always nonnegative.

Necessary conditions to obtain a sensitivity beyond the classical Fisher information~(\ref{eq:ChiclassicalF}) are that at least one covariance $\mathrm{Cov}(\hat{H},\hat{\Pi}_x)_{\hat{\rho}(\theta)}$ is nonzero and that $\hat{H}$ is not diagonal in $\hat{\boldsymbol{\Pi}}$. Even though a sensitivity above the classical Fisher information (for the projectors $\hat{\boldsymbol{\Pi}}$) can be achieved this way, the quantum Fisher information $F_Q[\hat{\rho},\hat{H}]$ always provides an upper sensitivity limit and we find the hierarchy
\begin{align}\label{eq:hierarchy}
\chi^{-2}_{\rm max}[\hat{\rho}(\theta),\hat{H},\hat{\boldsymbol{\Pi}}]\leq \chi^{-2}_{\rm max}[\hat{\rho}(\theta),\hat{H},\hat{\mathbf{H}}]\leq F_Q[\hat{\rho},\hat{H}],
\end{align}
where $F_Q[\hat{\rho},\hat{H}]=\max_{\hat{\boldsymbol{\Pi}}}F[\hat{\rho}(\theta),\hat{\boldsymbol{\Pi}}]$ is independent of $\theta$ \cite{BraunsteinPRL1994}. In other words, access to $\hat{H}$ permits us to overcome the classical Cram\'{e}r-Rao bound for the basis $\hat{\boldsymbol{\Pi}}$ but the sensitivity is of course always limited by the quantum Cram\'{e}r-Rao bound $(\Delta\theta_{\rm est})^2\geq (\Delta\theta_{\rm QCR})^2$ with $(\Delta\theta_{\rm QCR})^2=\min_{\hat{\boldsymbol{\Pi}}}(\Delta\theta_{\rm CR, \hat{\boldsymbol{\Pi}}})^2=(\mu F_Q[\hat{\rho},\hat{H}])^{-1}$. The proofs for Eqs.~(\ref{eq:chiQPiH})--(\ref{eq:hierarchy}) are given in Appendix~\ref{app:proofs}.

These results confirm our intuition that access to a larger family of measurement observables can only enhance the sensitivity. However, to provide high sensitivity, the measurement result should depend strongly on changes of the parameter $\theta$, whereas the generating Hamiltonian $\hat{H}$ is entirely insensitive [the commutator in Eq.~(\ref{eq:chi}) is zero if we measure $\hat{X}=\hat{H}$]. Therefore, the fact that the enhancement $E[\hat{\rho}(\theta),\hat{H},\hat{\boldsymbol{\Pi}}]$ is nonzero is not entirely evident. The example of an atomic clock in Sec.~\ref{sec:atomic} shows that this enhancement can indeed be significant and scale linearly with the total number of atoms.

\section{Optimal observable and implementation}
The maximum sensitivity~(\ref{eq:chiQPiH}) is achieved by the optimal observable (up to arbitrary constants that can be used to normalize the coefficients)
\begin{align}\label{eq:xopt}
\hat{X}_{\mathrm{opt}}=\hat{X}_{\mathrm{opt},0}+ab\left(\sum_{x=1}^{r}\frac{1}{p(x|\theta)}\mathrm{Cov}(\hat{H},\hat{\Pi}_x)_{\hat{\rho}(\theta)}\hat{\Pi}_x-\hat{H}\right).
\end{align}
A proof is provided in Appendix~\ref{app:Xopt}. The observable
\begin{align}\label{eq:xopt0}
\hat{X}_{\mathrm{opt},0}=\sum_{x=1}^r\left(\frac{\partial}{\partial \theta} \log p(x|\theta)\right)\hat{\Pi}_x
\end{align}
is optimal if only linear combinations of the $\hat{\boldsymbol{\Pi}}$ can be measured but $\hat{H}$ remains inaccessible \cite{GessnerPRL2019} as it achieves the maximum in Eq.~(\ref{eq:ChiclassicalF}) \cite{Kholevo,Frowis2015}. The observable $\hat{X}_{\mathrm{opt},0}$ is still optimal even if $\hat{H}$ could be measured when $\mathrm{Cov}(\hat{H},\hat{\Pi}_x)_{\hat{\rho}(\theta)}=0$ for all $x$, since in this case all other contributions to $\hat{X}_{\mathrm{opt}}$ vanish due to $b=0$. Both $\hat{X}_{\mathrm{opt}}$ and $\hat{X}_{\mathrm{opt},0}$ are defined at a fixed value of $\theta$.

We may wonder how the variance and the commutator part of the inverse parameter $\chi^{-2}$, Eq.~(\ref{eq:chi}), are affected by measuring contributions proportional to $\hat{H}$. Interestingly, it turns out that both observables~(\ref{eq:xopt}) and~(\ref{eq:xopt0}) have the property that numerator and the square root of the denominator of the squeezing coefficient~(\ref{eq:chi}) coincide and yield the maximum sensitivity~(\ref{eq:ChiclassicalF}). Specifically, if we use the definition~(\ref{eq:xopt}), it is straightforward to see that
\begin{align}
(\Delta \hat{X}_{\mathrm{opt}})^2_{\hat{\rho}(\theta)}=F[\hat{\rho}(\theta),\hat{\boldsymbol{\Pi}}]+E[\hat{\rho}(\theta),\hat{H},\hat{\boldsymbol{\Pi}}]
\end{align}
and
\begin{align}
-i\langle[\hat{X}_{\mathrm{opt}},\hat{H}]\rangle_{\hat{\rho}(\theta)}=F[\hat{\rho}(\theta),\hat{\boldsymbol{\Pi}}]+E[\hat{\rho}(\theta),\hat{H},\hat{\boldsymbol{\Pi}}].
\end{align}
The former follows from $\langle(\hat{X}_{\mathrm{opt}})^2\rangle_{\hat{\rho}(\theta)}=F[\hat{\rho}(\theta),\hat{\boldsymbol{\Pi}}]+(ab)^2\left(\langle\hat{H}^2\rangle_{\hat{\rho}(\theta)}-\sum_{x=1}^r\frac{1}{p(x|\theta)}\mathrm{Cov}(\hat{H},\hat{\Pi}_x)^2_{\hat{\rho}(\theta)}\right)$  
and $\langle\hat{X}_{\mathrm{opt}}\rangle_{\hat{\rho}(\theta)}^2=(ab)^2\langle\hat{H}\rangle_{\hat{\rho}(\theta)}^2$. Analogously, with Eq.~(\ref{eq:xopt0}) we obtain
\begin{align}
(\Delta \hat{X}_{\mathrm{opt},0})^2_{\hat{\rho}(\theta)}=F[\hat{\rho}(\theta),\hat{\boldsymbol{\Pi}}],
\end{align}
with $\langle\hat{X}_{\mathrm{opt},0}\rangle_{\hat{\rho}(\theta)}=0$, and
\begin{align}
-i\langle[\hat{X}_{\mathrm{opt},0},\hat{H}]\rangle_{\hat{\rho}(\theta)}=F[\hat{\rho}(\theta),\hat{\boldsymbol{\Pi}}].
\end{align}
Hence, both the commutator and the variance grow by the same amount $E[\hat{\rho}(\theta),\hat{H},\hat{\boldsymbol{\Pi}}]$ when measuring $\hat{X}_{\mathrm{opt}}$ instead of $\hat{X}_{\mathrm{opt},0}$. Because the metrological sensitivity~(\ref{eq:chi}) scales with the square of the commutator, the enhancement $E[\hat{\rho}(\theta),\hat{H},\hat{\boldsymbol{\Pi}}]$ is directly added to the sensitivity.

The implementation of the improved scheme is based on the estimation of $\theta$ from measurements of the average value of $\hat{X}_{\rm opt}$, which in turn is a linear combination of the type 
\begin{align}\label{eq:obsX}
\hat{X}=\sum_{x=1}^rc_x\hat{\Pi}_x+c_H\hat{H}
\end{align}
with real coefficients $c_1,\dots,c_r$ and $c_H$. The expectation value is given by $\langle\hat{X}\rangle_{\hat{\rho}(\theta)}=\sum_{x=1}^rc_xp(x|\theta)+c_H\langle\hat{H}\rangle_{\hat{\rho}(\theta)}$. Notice that $\langle \hat{H}\rangle_{\hat{\rho}(\theta)}= \langle \hat{H}\rangle_{\hat{\rho}}$ is independent of $\theta$, and therefore a property of the initial state. Assuming that this additional piece of \textit{a priori} information about the initial state is available before the experiment, the expectation value of any observable $\hat{X}$ of the type~(\ref{eq:obsX}) can be obtained with access to the basis $\hat{\boldsymbol{\Pi}}$, which provides the part of $\langle\hat{X}\rangle_{\hat{\rho}(\theta)}$ that depends on the $p(x|\theta)$. It is reasonable to assume knowledge of $\langle\hat{H}\rangle_{\hat{\rho}}$ in experimentally relevant cases, since the initial state $\hat{\rho}$ and the phase-imprinting generator are usually well known in phase-estimation experiments. The considered scenario of a single unknown parameter of fixed value assumes that all other parameters with influence on the measurement outcomes are known \cite{Yan2018}. 

The method suggested above reconstructs the average value of $\hat{X}$ by combining the \textit{a priori} information on energy with the measurement results in $\hat{\boldsymbol{\Pi}}$. Knowledge of the energy thus avoids the need for a direct measurement in the basis of $\hat{X}$. However, the obtained measurement results are equivalent only on average, while their statistics are completely different. It is therefore important to notice that the proposed scheme is based on the method of moments and requires only knowledge of the average value of $\hat{X}$. If access to the full counting statistics was available, an equally optimal estimation strategy would be given by a maximum likelihood estimation. To obtain the counting statistics of $\hat{X}$, however, the projectors onto its eigenvalues must be measured. This would in general constitute a challenging task since the proposed scheme only provides an advantage when $\hat{H}$, and consequently $\hat{X}$, is not diagonal in $\hat{\boldsymbol{\Pi}}$.

Let us finally remark that the exact coefficients of the optimal observable [Eqs.~(\ref{eq:xopt}) and~(\ref{eq:xopt0})] depend on the value of the phase $\theta$, which is not known in realistic settings. Even though an implementation of the optimal observable may not be practical in an experiment, it is important to identify the ultimate precision limit of the proposed strategy. Moreover, the method discussed above is not limited to the optimal observable and can be implemented with arbitrary coefficients. Any realistic implementation gives rise to a lower bound for the optimal sensitivity that is studied here. More generally, our results show that whenever the additional information about energy is taken explicitly into consideration, the Cram\'{e}r-Rao bound associated with the measurement basis no longer poses a limit to the achievable sensitivity.

\section{Atomic clock with oversqueezed spin states}\label{sec:atomic}

\begin{figure*}[tb]
\centering
\includegraphics[width=.98\textwidth]{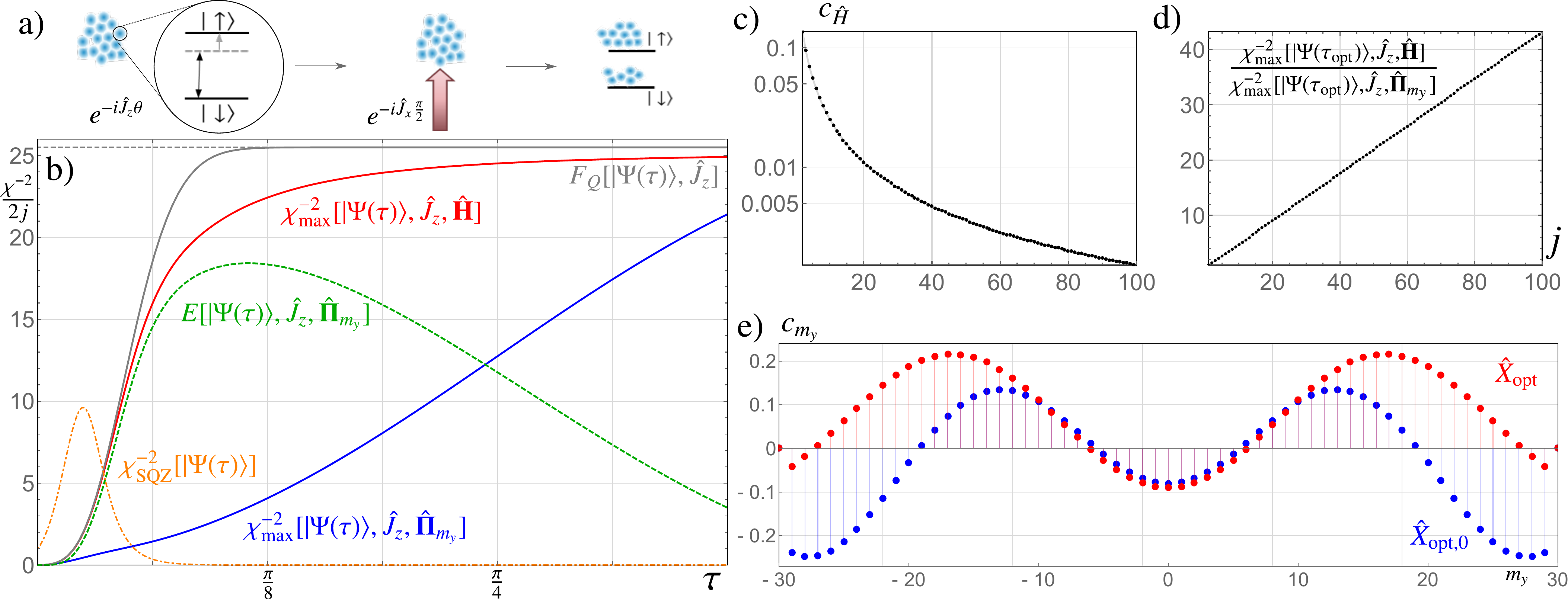}
\caption{(a) The clock measurement consists in free evolution with the Hamiltonian $\hat{H}=\hat{J}_z$, followed by a $\pi/2$-rotation around the $x$ axis and a measurement of the number of atoms in the two internal states. (b) Sensitivity limits rescaled by the shot noise $N=2j$ for the state $|\Psi(t)\rangle$ with $j=25$. We compare the maximal measurement sensitivity~(\ref{eq:ChiclassicalF}) obtained by the eigenstates of $\hat{J}_y$ (blue line) to the sensitivity~(\ref{eq:chiQPiH}) enhanced by additional Hamiltonian measurements (red line). The enhancement is shown as the green dashed line. The upper sensitivity bound is given by the quantum Fisher information [gray line, see Eq.~(\ref{eq:hierarchy})], which reaches the value $j(2j+1)$ (gray dashed line). For comparison, we show the spin squeezing coefficient (orange dot-dashed line). The maximal enhancement is found in oversqueezed states around $\tau_{\rm opt}\simeq 0.94/\sqrt{j}$. (c) Normalized coefficient $c_{\hat{H}}$, expressing the contribution $\hat{H}$ as a function of $j$ at the point $\tau_{\rm opt}$ of maximal enhancement. (d) Despite the near-vanishing contribution of $\hat{H}$, the relative gain at $\tau_{\rm opt}$, expressed by the ratio of Eqs.~(\ref{eq:ChiclassicalF}) and~(\ref{eq:chiQPiH}), increases linearly with $j=N/2$. (e) Coefficients $c_{m_y}$ of the optimal measurement observables with ($\hat{X}_{\rm opt}$, red dots) and without ($\hat{X}_{\rm opt}^{(0)}$, blue dots) measurement access to $\hat{H}$ for $j=100$. We have $c_{m_y}=0$ for all $|m_y|\geq 30$ for both $\hat{X}_{\rm opt}$ and $\hat{X}_{\rm opt,0}$. The normalized contribution of $\hat{H}$ to $\hat{X}_{\rm opt}$ is $c_{\hat{H}}\lesssim 2\times 10^{-3}$.}
\label{fig:1}
\end{figure*}

We now apply these results to the example of an atomic clock, composed of $N$ spin-$1/2$ particles, described by collective spin operators $\hat{J}_{\alpha}=\frac{1}{2}\sum_{i=1}^N\hat{\sigma}_{\alpha,i}$, where $\alpha=x,y,z$ and $\hat{\sigma}_{\alpha,i}$ are local Pauli matrices for the $i$th spin. The basic clock operation consists in a precise estimation of the atomic resonance frequency between ground- and excited state by Ramsey spectroscopy~\cite{SchmidtRMP}. The time evolution generated by the Hamiltonian $\hat{J}_z$ imprints the phase parameter $\theta$ that is directly proportional to this resonance frequency. Let us further assume that after the phase imprinting, the statistics of the observable $\hat{J}_y$ can be measured by realizing a $\pi/2$ rotation of the spins around the $x$ axis followed by a measurement of the number of atoms in the ground- or excited states [see Fig.~\ref{fig:1}(a)]. 

Quantum-enhanced sensitivities can be achieved with spin states generated by the nonlinear one-axis-twisting evolution \cite{KitagawaUeda,RMP}. By subjecting a spin-coherent state $|j,j\rangle_z$ with all spins polarized along the $z$ axis~\cite{MandelWolf} to a nonlinear evolution generated by $\hat{J}_y^2$, we obtain the states $|\Psi(\tau)\rangle=e^{-i\hat{J}_y^2\tau}|j,j\rangle_z$ and $j=N/2$ is the total spin length. After short times $\tau$, these states are still well characterized by Gaussian measurements, i.e., mean values and variances of collective spin observables, and their sensitivity as well as their entanglement is captured by the spin squeezing parameter~\cite{Wineland,Sorensen,SMPRL01,Toth,Ma,Appelaniz,Sinatra,RMP}. As $\tau$ increases, these states become more sensitive for the atomic clock measurement considered above, until reaching a maximally sensitive NOON state at $\tau=\pi/2$. We focus on non-Gaussian quantum states that are generated at longer time scales than spin-squeezed states but shorter than those required to reach the NOON state. Several techniques can lead to a precision enhancement in this scenario. Standard methods consist in shifting the value of $\theta$ to an optimal value or implementing additional spin rotations to change the directions of the collective spin operators that determine the generator or the measurement~\cite{RMP}. Here, we do not make use of these methods and instead focus on the sensitivity gain, Eq.~(\ref{eq:chiQPiH}), that can be provided by additional measurements of the generating Hamiltonian $\hat{J}_z$ in the scenario where both $\theta$ and the original observable $\hat{J}_y$ are fixed. 

The sensitivity limits, rescaled by the shot-noise level $N=2j$, are represented in Fig.~\ref{fig:1}(b) for the states $|\Psi(\tau)\rangle$. The blue line shows the optimal sensitivity~(\ref{eq:ChiclassicalF}) for measurements that are limited to observables which are diagonal in the basis spanned by the eigenstates $|j,-j\rangle_y,|j,-j+1\rangle_y,\dots,|j,j-1\rangle_y,|j,j\rangle_y$ of $\hat{J}_y$. Here, the $N+1$ projectors $\hat{\Pi}_{m_y}=|j,m_y\rangle_y\langle j,m_y|_y$ with $m_y=-j,\dots,j$ represent the family of accessible operators $\hat{\boldsymbol{\Pi}}_{m_y}$. If in addition to $\hat{\boldsymbol{\Pi}}_{m_y}$ also the phase-imprinting generator $\hat{J}_z$ can be measured, we obtain the maximally achievable sensitivity~(\ref{eq:chiQPiH}) displayed by the red line. The green dashed line represents the enhancement $E[|\Psi(\tau)\rangle,\hat{J}_z,\hat{\boldsymbol{\Pi}}_{m_y}]$. The upper bound, provided by the quantum Fisher information $F_Q[|\Psi(\tau)\rangle,\hat{J}_z]$, is displayed by the gray line and we observe the hierarchy of bounds~(\ref{eq:hierarchy}). As lower bounds on $F_Q[|\Psi(\tau)\rangle,\hat{J}_z]$, the sensitivities in this plot can be compared to separability bounds on the quantum Fisher information (without restricting to pure states). For instance, any indication of $\chi^{-2}/(2j)>k$ indicates at least $k$ entangled particles \cite{HyllusPRA2012}. Hence, the sensitivity enhancement directly provides an improved entanglement witnesses.

For comparison, we show the spin squeezing coefficient \cite{Wineland,RMP}, which can be obtained by maximizing Eq.~(\ref{eq:chi}) over all measurement observables and Hamiltonians $\hat{J}_{\mathbf{n}}=\mathbf{n}\cdot\hat{\mathbf{J}}$ that can be written as linear combinations of $\hat{\mathbf{J}}=(\hat{J}_x,\hat{J}_y,\hat{J}_z)$ as $\chi^{-2}_{\rm SQZ}[|\Psi(\tau)\rangle]=\max_{\mathbf{n}}\chi^{-2}_{\max}[|\Psi(\tau)\rangle,\hat{J}_{\mathbf{n}},\hat{\mathbf{J}}]$ \cite{GessnerPRL2019}. The decay of $\chi^{-2}_{\rm SQZ}[|\Psi(\tau)\rangle]$ indicates the loss of Gaussianity as the state becomes oversqueezed and is no longer well characterized by the covariances of $\hat{\mathbf{J}}$ \cite{RMP}. We observe the maximal enhancement $E[|\Psi(\tau)\rangle,\hat{J}_z,\hat{\boldsymbol{\Pi}}_{m_y}]$ at $\tau_{\mathrm{opt}}\simeq 0.94/\sqrt{j}$ when spin squeezing is almost entirely lost.

We introduce the coefficients $c_{\hat{H}}$ and $c_{m_y}$ to represent measurement observables as $\hat{X}=c_{\hat{H}}\hat{H}+\sum_{m_y=-j}^jc_{m_y}\hat{\Pi}_{m_y}$ with the normalization $|c_{\hat{H}}|^2+\sum_{m_y=-j}^j|c_{m_y}|^2=1$. The contribution $c_{\hat{H}}$ is always zero for $\hat{X}_{\rm opt,0}$. For $\hat{X}_{\rm opt}$, we find that the coefficient $c_{\hat{H}}$ tends toward zero at $\tau_{\rm opt}$ as $j$ increases; see Fig.~\ref{fig:1}(c). Nevertheless, the relative sensitivity gain obtained by measuring $\hat{X}_{\mathrm{opt}}$ instead of $\hat{X}_{\mathrm{opt},0}$ can be quite significant as is shown by the linear scaling with the number of particles in Fig.~\ref{fig:1}(d). This shows that the metrological sensitivity may depend strongly on tiny changes of the measurement observable.

Being able to measure $\hat{H}$ can have dramatic influence on the weight of different projectors [cf. Eqs.~(\ref{eq:xopt}) and~(\ref{eq:xopt0})], as is shown for $j=100$ in Fig.~\ref{fig:1}(e). The contribution of $\hat{H}$ to $\hat{X}_{\rm opt}$ is very small, $|c_{\hat{H}}|\lesssim 2\times 10^{-3}$, in this case. As negligibly small as it may seem, the contribution of $\hat{H}$ cannot be ignored. By removing the Hamiltonian part from Eq.~(\ref{eq:xopt}), we obtain the observable $\hat{X}_{\mathrm{opt}}+ab\hat{H}$, which can be implemented without access to $\hat{H}$. However, according to Eqs.~(\ref{eq:ChiclassicalF}) and~(\ref{eq:xopt0}) this yields a suboptimal measurement and a sensitivity below $F[|\Psi(\tau)\rangle,\hat{\boldsymbol{\Pi}}_{m_y}]$. In other words, if $c_{\hat{H}}$ is set to zero in $\hat{X}_{\rm opt}$, the sensitivity in Fig.~\ref{fig:1}(b) drops from the red line to a value below the blue line. It can be easily verified that even though the observable $\hat{X}_{\mathrm{opt}}+ab\hat{H}$ has the same gradient [denominator in Eq.~(\ref{eq:chi})] as $\hat{X}_{\mathrm{opt}}$, its variance [the numerator in Eq.~(\ref{eq:chi})] is much larger, which leads to a drastic reduction in sensitivity. 

\section{Conclusions}
In conclusion, we have shown how the precision of a phase measurement in a suboptimal basis can be improved by knowledge of the initial state's expectation value for the Hamiltonian $\hat{H}$ that generates the phase shift. This is despite the fact that $\langle\hat{H}\rangle_{\hat{\rho}}$ itself is entirely insensitive of the phase. If the \textit{a priori} information on energy is utilized in an optimal way, the proposed method is equivalent to the measurement of an optimal linear combination [Eq.~(\ref{eq:xopt})] of the basis projectors and $\hat{H}$. The classical Cram\'er-Rao bound associated with the accessible basis no longer poses a limit on the achievable sensitivity [Eq.~(\ref{eq:chiQPiH})].

For the example of an atomic clock we found a sensitivity gain that scales linearly with the number of atoms. Concerning the optimal observable, access to $\hat{H}$ mostly entails a significant shift of the contribution of the projectors that were already accessible without access to $\hat{H}$, while the contribution of $\hat{H}$ itself is tiny. Nevertheless, the contribution of $\hat{H}$ to the optimal observable cannot be neglected, as it would lead to a drastic reduction of the sensitivity. This hints at a discontinuity of the achievable sensitivity when a part of the available information disappears. Discontinuous behavior of the Fisher information was recently studied in Refs.~\cite{Safranek17,Seveso19} but a possible relation to these observations remains open for future investigations. More generally, this shows that the sensitivity of phase estimation experiments based on the widely used method of moments can depend strongly on the precise implementation of the measurement observable. 

\section*{Acknowledgments} I would like to thank A. Smerzi and L. Pezz\`{e} for countless enlightening discussions on quantum metrology over the past years. This work was funded by the ENS-ICFP LabEx Grant No. ANR-10-LABX-0010/ANR-10-IDEX-0001-02 PSL*.

\appendix
\section{Derivation of the main results}\label{app:proofs}
\subsection{Proof of Eq.~(\ref{eq:chiQPiH})}
Let us first derive the main result of this article, Eq.~(\ref{eq:chiQPiH}). The analytical optimization over the measurement observable $\hat{X}$ in Eq.~(\ref{eq:chiQPiH}) over an arbitrary family $\hat{\mathbf{H}}=(\hat{H}_1,\dots,\hat{H}_L)$ of accessible operators is given by \cite{GessnerPRL2019}
\begin{align}\label{eq:thm}
\max_{\hat{X}\in\mathrm{span}(\hat{\mathbf{H}})}\chi^{-2}[\hat{\rho}(\theta),\hat{H},\hat{X}]=\mathbf{n}^T\mathbf{M}[\hat{\rho}(\theta),\hat{\mathbf{H}}]\mathbf{n},
\end{align}
where $\hat{H}=\mathbf{n}^T\mathbf{H}$ with $\mathbf{n}\in\mathbb{R}^L$. The moment matrix
\begin{align}\label{eq:momentmatrixNOX}
\mathbf{M}[\hat{\rho},\hat{\mathbf{H}}]=\mathbf{C}[\hat{\rho},\hat{\mathbf{H}}]^T\boldsymbol{\Gamma}[\hat{\rho},\hat{\mathbf{H}}]^{-1}\mathbf{C}[\hat{\rho},\hat{\mathbf{H}}]
\end{align}
is composed of the covariance matrix $(\boldsymbol{\Gamma}[\hat{\rho},\hat{\mathbf{H}}])_{kl}=\mathrm{Cov}(\hat{H}_k,\hat{H}_l)_{\hat{\rho}}=\frac{1}{2}\langle \hat{H}_k\hat{H}_l+\hat{H}_l\hat{H}_k\rangle_{\hat{\rho}}-\langle\hat{H}_k\rangle_{\hat{\rho}}\langle\hat{H}_l\rangle_{\hat{\rho}}$ and the commutator matrix $(\mathbf{C}[\hat{\rho},\hat{\mathbf{H}}])_{kl}=-i\langle[\hat{H}_k,\hat{H}_l]\rangle_{\hat{\rho}}$.

In the following, we analytically determine the right-hand side of Eq.~(\ref{eq:thm}) for the case described in Eq.~(\ref{eq:chiQPiH}). Notice that if we include a complete, orthonormal family of operators $\hat{\boldsymbol{\Pi}}$ in the set $\hat{\mathbf{H}}$ of accessible operators, the covariance matrix $\boldsymbol{\Gamma}[\hat{\rho}(\theta),\hat{\mathbf{H}}]$ becomes singular, since the completeness relation $\sum_{x}\hat{\Pi}_x=\hat{\mathbb{I}}$ renders the information provided by the measurement of a single projector dependent on the results of all other projectors. To avoid the singularity of $\boldsymbol{\Gamma}[\hat{\rho}(\theta),\hat{\mathbf{H}}]$, we thus remove one of the projectors from the complete set $\hat{\mathbf{\Pi}}=\{\hat{\Pi}_1,\dots,\hat{\Pi}_r\}$ to obtain the set $\hat{\mathbf{\Pi}}_{\{r\}}=\{\hat{\Pi}_1,\dots,\hat{\Pi}_{r-1}\}$ with $\sum_{x=1}^{r-1}\hat{\Pi}_x=\hat{\mathbb{I}}-\hat{\Pi}_r$. The set of accessible operators $\hat{\mathbf{H}}=(\hat{H},\hat{\Pi}_1,\dots,\hat{\Pi}_{r-1})$ is composed of the Hamiltonian $\hat{H}$ and the projectors $\hat{\mathbf{\Pi}}_{\{r\}}$. 

To apply the result~(\ref{eq:thm}), we first identify the moment matrix $\mathbf{M}[\hat{\rho}(\theta),\hat{\mathbf{H}}]$, which in turn is composed of the covariance and commutator matrices. We obtain the covariance matrix
\begin{align}\label{eq:gammaX}
\boldsymbol{\Gamma}[\hat{\rho}(\theta),\hat{\mathbf{H}}]=\begin{pmatrix}(\Delta \hat{H})^2_{\hat{\rho}(\theta)} & \boldsymbol{\gamma}^T\\
\boldsymbol{\gamma} & \boldsymbol{\Gamma}[\hat{\rho}(\theta),\hat{\mathbf{\Pi}}_{\{r\}}]
\end{pmatrix},
\end{align}
where $\boldsymbol{\gamma}$ is a vector with elements $\gamma_x=\mathrm{Cov}(\hat{H},\hat{\Pi}_x)_{\hat{\rho}(\theta)}$ for $x=1,\dots,r-1$, and $\boldsymbol{\Gamma}[\hat{\rho}(\theta),\hat{\mathbf{\Pi}}_{\{r\}}]$ is the $(r-1)\times(r-1)$ covariance matrix of the projectors $\hat{\mathbf{\Pi}}_{\{r\}}$. Since all elements in $\hat{\boldsymbol{\Pi}}$ are orthogonal and thus commute, the commutator matrix reads
\begin{align}\label{eq:Cproj}
\mathbf{C}[\hat{\rho}(\theta),\hat{\mathbf{H}}]=\begin{pmatrix}0& -\mathbf{d}^T\\
\mathbf{d} & \mathbf{0}
\end{pmatrix},
\end{align}
where the vector $\mathbf{d}$ has elements $d_x=-i\langle [\hat{\Pi}_x,\hat{H}]\rangle_{\hat{\rho}(\theta)}=\frac{\partial p(x|\theta)}{\partial \theta}$ with $p(x|\theta)$ for $x=1,\dots,r-1$, and $\mathbf{0}$ is a $(r-1)\times(r-1)$ zero matrix. We note that it is essential for the $\hat{\Pi}_x$ to not commute with $\hat{H}$ in order to obtain a useful bound. This was expected since otherwise $\hat{H}$ would have already been included in the set of observables that can be constructed as linear combinations of the elements of $\hat{\boldsymbol{\Pi}}$.

Recall from Eq.~(\ref{eq:momentmatrixNOX}) that $\mathbf{M}[\hat{\rho}(\theta),\hat{\mathbf{H}}]$ is a function of the inverse matrix of $\boldsymbol{\Gamma}[\hat{\rho}(\theta),\hat{\mathbf{H}}]$. Making use of its block structure, the inverse of~(\ref{eq:gammaX}) can be written as \cite{Lu2002}
\begin{align}\label{eq:invGammaX}
\boldsymbol{\Gamma}[\hat{\rho}(\theta),\hat{\mathbf{H}}]^{-1}=\begin{pmatrix}a & -a\mathbf{w}^T\\
-a\mathbf{w} & \quad\boldsymbol{\Gamma}[\hat{\rho}(\theta),\hat{\mathbf{\Pi}}_{\{r\}}]^{-1}+a\mathbf{w}\mathbf{w}^T
\end{pmatrix},
\end{align}
where $a=\{(\Delta\hat{H})^2_{\hat{\rho}(\theta)}-\boldsymbol{\gamma}^T\boldsymbol{\Gamma}[\hat{\rho}(\theta),\hat{\mathbf{\Pi}}_{\{r\}}]^{-1}\boldsymbol{\gamma}\}^{-1}$ and $\mathbf{w}=\boldsymbol{\Gamma}[\hat{\rho}(\theta),\hat{\mathbf{\Pi}}_{\{r\}}]^{-1}\boldsymbol{\gamma}$. 

We now take a closer look at the elements of the covariance matrix, which read $(\boldsymbol{\Gamma}[\hat{\rho}(\theta),\hat{\mathbf{\Pi}}_{\{r\}}])_{xx'}=\mathrm{Cov}(\hat{\Pi}_x,\hat{\Pi}_{x'})_{\hat{\rho}(\theta)}=\delta_{xx'}p(x|\theta)-p(x|\theta)p(x'|\theta)$. Hence, we obtain $\boldsymbol{\Gamma}[\hat{\rho}(\theta),\hat{\mathbf{\Pi}}_{\{r\}}]=\mathbf{P}_{\theta}-\mathbf{p}_{\theta}\mathbf{p}_{\theta}^T$, where $\mathbf{P}_{\theta}=\mathrm{diag}(p(1|\theta),\dots,p(r-1|\theta))$ is a diagonal matrix whose diagonal elements define the vector $\mathbf{p}^T_{\theta}=(p(1|\theta),\dots,p(r-1|\theta))$ and $\mathbf{p}_{\theta}\mathbf{p}_{\theta}^T$ is a rank-1 matrix. The matrix inverse can be determined using the result given in Ref.~\cite{Miller81} and reads $\boldsymbol{\Gamma}[\hat{\rho}(\theta),\hat{\mathbf{\Pi}}_{\{r\}}]^{-1}=\mathbf{P}_{\theta}^{-1}+\frac{1}{1-\mathbf{p}_{\theta}^T\mathbf{P}_{\theta}^{-1}\mathbf{p}_{\theta}}\mathbf{P}_{\theta}^{-1}\mathbf{p}_{\theta}\mathbf{p}_{\theta}^T\mathbf{P}_{\theta}^{-1}$, where $\mathbf{P}_{\theta}^{-1}=\mathrm{diag}(\frac{1}{p(1|\theta)},\dots,\frac{1}{p(r-1|\theta)})$ and we assume that $p(x|\theta)>0$ for all $x=1,\dots,r$. The result can be further simplified by noticing that $\mathbf{e}=\mathbf{P}_{\theta}^{-1}\mathbf{p}_{\theta}=(1,\dots,1)^T$ and $\mathbf{p}_{\theta}^T\mathbf{P}_{\theta}^{-1}\mathbf{p}_{\theta}=\sum_{x=1}^{r-1}p(x|\theta)=1-p(r|\theta)$, finally leading to
\begin{align}\label{eq:invGammaPi}
\boldsymbol{\Gamma}[\hat{\rho}(\theta),\hat{\mathbf{\Pi}}_{\{r\}}]^{-1}=\mathbf{P}_{\theta}^{-1}+\frac{1}{p(r|\theta)}\mathbf{e}\mathbf{e}^T.
\end{align}

Inserting Eq.~(\ref{eq:invGammaPi}) back into Eq.~(\ref{eq:invGammaX}), we obtain
\begin{align}\label{eq:s}
a^{-1}=(\Delta\hat{H})^2_{\hat{\rho}(\theta)}-\sum_{x=1}^r\frac{1}{p(x|\theta)}\mathrm{Cov}(\hat{H},\hat{\Pi}_x)^2_{\hat{\rho}(\theta)}
\end{align}
and the vector $\mathbf{w}=\mathbf{P}_{\theta}^{-1}\boldsymbol{\gamma}+\frac{1}{p(r|\theta)}\mathbf{e}(\mathbf{e}^T\boldsymbol{\gamma})$ has elements $w_x=\frac{1}{p(x|\theta)}\mathrm{Cov}(\hat{H},\hat{\Pi}_x)_{\hat{\rho}(\theta)}-\frac{1}{p(r|\theta)}\mathrm{Cov}(\hat{H},\hat{\Pi}_r)_{\hat{\rho}(\theta)}$ for $x=1,\dots,r-1$. Here, we made use of the identities $\boldsymbol{\gamma}^T\boldsymbol{\Gamma}[\hat{\rho}(\theta),\hat{\mathbf{\Pi}}_{\{r\}}]^{-1}\boldsymbol{\gamma}=\boldsymbol{\gamma}^T\mathbf{P}_{\theta}^{-1}\boldsymbol{\gamma}+\frac{1}{p(r|\theta)}(\mathbf{e}^T\boldsymbol{\gamma})^2=\sum_{x=1}^{r}\frac{1}{p(x|\theta)}\mathrm{Cov}(\hat{H},\hat{\Pi}_x)^2_{\hat{\rho}(\theta)}$, due to $\boldsymbol{\gamma}^T\mathbf{P}_{\theta}^{-1}\boldsymbol{\gamma}=\sum_{x=1}^{r-1}\frac{1}{p(x|\theta)}\mathrm{Cov}(\hat{H},\hat{\Pi}_x)^2_{\hat{\rho}(\theta)}$ and $\mathbf{e}^T\boldsymbol{\gamma}=\sum_{x=1}^{r-1}\mathrm{Cov}(\hat{H},\hat{\Pi}_x)_{\hat{\rho}(\theta)}=\mathrm{Cov}(\hat{H},\sum_{x=1}^{r-1}\hat{\Pi}_x)_{\hat{\rho}(\theta)}=-\mathrm{Cov}(\hat{H},\hat{\Pi}_r)_{\hat{\rho}(\theta)}$.

Next, we obtain
\begin{align}\label{eq:GammaC}
&\quad\boldsymbol{\Gamma}[\hat{\rho}(\theta),\hat{\mathbf{H}}]^{-1}\mathbf{C}[\hat{\rho}(\theta),\hat{\mathbf{H}}]\notag\\&=\begin{pmatrix}-a\mathbf{w}^T\mathbf{d} & -a\mathbf{d}^T\\
\boldsymbol{\Gamma}[\hat{\rho}(\theta),\hat{\mathbf{\Pi}}_{\{r\}}]^{-1}\mathbf{d}+a\mathbf{w}(\mathbf{w}^T\mathbf{d})&\:a\mathbf{w}\mathbf{d}^T
\end{pmatrix},
\end{align}
where from $\sum_{x=1}^r\frac{\partial}{\partial \theta} p(x|\theta)=\frac{\partial}{\partial \theta}1=0$ follows that
\begin{align}\label{eq:wTd}
\mathbf{w}^T\mathbf{d}=\sum_{x=1}^r\mathrm{Cov}(\hat{H},\hat{\Pi}_x)_{\hat{\rho}(\theta)}\left(\frac{\partial}{\partial \theta} \log p(x|\theta)\right)
\end{align}
and $\boldsymbol{\Gamma}[\hat{\rho}(\theta),\hat{\mathbf{\Pi}}_{\{r\}}]^{-1}\mathbf{d}=(\frac{\partial}{\partial \theta} \log \frac{p(1|\theta)}{p(r|\theta)},\dots,\frac{\partial}{\partial \theta} \log \frac{p(r-1|\theta)}{p(r|\theta)})^T$. Finally, the moment matrix~(\ref{eq:momentmatrixNOX}) reads
\begin{align}
\mathbf{M}[\hat{\rho}(\theta),\hat{\mathbf{H}}]=\begin{pmatrix}\mathbf{d}^T\boldsymbol{\Gamma}[\hat{\rho}(\theta),\hat{\mathbf{\Pi}}_{\{r\}}]^{-1}\mathbf{d}+a(\mathbf{w}^T\mathbf{d})^2 &\: a(\mathbf{w}^T\mathbf{d})\mathbf{d}\\
a(\mathbf{w}^T\mathbf{d})\mathbf{d} & a\mathbf{d}\mathbf{d}^T
\end{pmatrix},\notag
\end{align}
where $\mathbf{d}^T\boldsymbol{\Gamma}[\hat{\rho}(\theta),\hat{\mathbf{\Pi}}_{\{r\}}]^{-1}\mathbf{d}=\sum_{x=1}^rp(x|\theta)(\partial \log p(x|\theta)/\partial\theta)^2=F[\hat{\rho}(\theta),\hat{X}]$ is the Fisher information. Now applying Eq.~(\ref{eq:thm}) for $\hat{H}=\mathbf{e}_1^T\hat{\mathbf{H}}$, where $\mathbf{e}_1=(1,0,\dots,0)^T\in\mathbb{R}^r$ is the first unit vector, we obtain the maximal sensitivity
\begin{align}
\max_{\hat{X}\in\mathrm{span}(\hat{\mathbf{H}})}\chi^{-2}[\hat{\rho}(\theta),\hat{H},\hat{X}]&=\mathbf{e}_1^T\mathbf{M}[\hat{\rho}(\theta),\hat{\mathbf{H}}]\mathbf{e}_1\notag\\&=F[\hat{\rho}(\theta),\hat{\boldsymbol{\Pi}}]+a(\mathbf{w}^T\mathbf{d})^2.\notag
\end{align}
This completes the proof of Eq.~(\ref{eq:chiQPiH}) with $E[\hat{\rho}(\theta),\hat{H},\hat{\boldsymbol{\Pi}}]=ab^2$, where $a$ and $b=\mathbf{w}^T\mathbf{d}$ are defined in Eqs.~(\ref{eq:s}) and~(\ref{eq:wTd}), respectively.

\subsection{Proof of the bounds~(\ref{eq:hierarchy})}
The hierarchy~(\ref{eq:hierarchy}) expresses that the optimized sensitivity that is achieved by adding energy measurements to the measurements of the observable $\hat{X}$ lies between the classical and quantum Fisher information. For the upper bound, see Ref.~\cite{GessnerPRL2019}. The lower bound holds since $a(\mathbf{w}^T\mathbf{d})^2\geq 0$, due to $a\geq 0$. This in turn follows from $a^{-1}=(\Delta\hat{H})^2_{\hat{\rho}(\theta)}-\sum_{x=1}^{r}\frac{1}{p(x|\theta)}\mathrm{Cov}(\hat{H},\hat{\Pi}_x)^2_{\hat{\rho}(\theta)}=\langle\hat{H}^2\rangle_{\hat{\rho}(\theta)}-\langle\hat{H}\rangle_{\hat{\rho}(\theta)}^2-\sum_{x=1}^r\left[\frac{1}{2}(\langle\hat{H}\hat{\Pi}_x+\hat{\Pi}_x\hat{H}\rangle_{\hat{\rho}(\theta)})-\langle\hat{H}\rangle_{\hat{\rho}(\theta)}\langle\hat{\Pi}_x\rangle_{\hat{\rho}(\theta)}\right]^2/\langle\hat{\Pi}_x\rangle_{\hat{\rho}(\theta)}=\langle\hat{H}^2\rangle_{\hat{\rho}(\theta)}-\sum_{x=1}^r\frac{1}{4}(\langle\hat{H}\hat{\Pi}_x+\hat{\Pi}_x\hat{H}\rangle_{\hat{\rho}(\theta)})^2/\langle\hat{\Pi}_x\rangle_{\hat{\rho}(\theta)}\geq \langle\hat{H}^2\rangle_{\hat{\rho}(\theta)}-\sum_{x=1}^r|\langle\hat{\Pi}_x\hat{H}\rangle_{\hat{\rho}(\theta)}|^2/\langle\hat{\Pi}_x\rangle_{\hat{\rho}(\theta)}$, where we used that $\frac{1}{4}(\langle\hat{H}\hat{\Pi}_x+\hat{\Pi}_x\hat{H}\rangle_{\hat{\rho}(\theta)})^2=\mathrm{Re}(\langle\hat{\Pi}_x\hat{H}\rangle_{\hat{\rho}(\theta)})^2\leq|\langle\hat{\Pi}_x\hat{H}\rangle_{\hat{\rho}(\theta)}|^2$. Using the Cauchy-Schwarz inequality $(\mathrm{Tr}\hat{A}^{\dagger}\hat{B})^2\leq (\mathrm{Tr}\hat{A}^{\dagger}\hat{A})(\mathrm{Tr}\hat{B}^{\dagger}\hat{B})$ with $\hat{A}=\hat{\Pi}_x\sqrt{\hat{\rho}(\theta)}$ and $\hat{B}=\hat{\Pi}_x\hat{H}\sqrt{\rho(\theta)}$, we obtain $\langle\hat{H}\hat{\Pi}_x\hat{H}\rangle_{\hat{\rho}(\theta)}\geq |\langle\hat{\Pi}_x\hat{H}\rangle_{\hat{\rho}(\theta)}|^2/\langle\hat{\Pi}_x\rangle_{\hat{\rho}(\theta)}$. Summation over $x$ on both sides now implies that $a\geq 0$.

\section{Optimal observable}\label{app:Xopt}
It was shown in Ref.~\cite{GessnerPRL2019} that the maximum sensitivity~(\ref{eq:thm}) is achieved by the optimal observable
\begin{align}\label{eq:optobs}
\hat{X}_{\rm opt}=\mathbf{m}^T\hat{\mathbf{H}}=\alpha\mathbf{n}^T\mathbf{C}[\hat{\rho},\hat{\mathbf{H}}]^T\boldsymbol{\Gamma}[\hat{\rho},\hat{\mathbf{H}}]^{-1}\hat{\mathbf{H}},
\end{align}
with some normalization constant $\alpha\in\mathbb{R}$. By applying Eq.~(\ref{eq:optobs}) to the case considered here with $\mathbf{n}=\mathbf{e}_1$, Eq.~(\ref{eq:GammaC}) and $\mathbf{w}^T \hat{\boldsymbol{\Pi}}_{\{r\}}=\sum_{x=1}^{r-1}[\frac{1}{p(x|\theta)}\mathrm{Cov}(\hat{H},\hat{\Pi}_x)_{\hat{\rho}(\theta)}-\frac{1}{p(r|\theta)}\mathrm{Cov}(\hat{H},\hat{\Pi}_r)_{\hat{\rho}(\theta)}]\hat{\Pi}_x=\sum_{x=1}^{r}\frac{1}{p(x|\theta)}\mathrm{Cov}(\hat{H},\hat{\Pi}_x)_{\hat{\rho}(\theta)}\hat{\Pi}_x+\alpha\hat{\mathbb{I}}$, we find up to irrelevant constants and normalization factors the optimal observable given in Eq.~(\ref{eq:xopt}). Notice further that $\hat{X}_{\mathrm{opt},0}=\sum_{x=1}^{r-1}[\frac{\partial}{\partial \theta} \log p(x|\theta)] \hat{\Pi}_x-(\hat{\mathbb{I}}-\hat{\Pi}_r)[\frac{\partial}{\partial \theta} \log p(r|\theta)]+\alpha'\hat{\mathbb{I}}=(\boldsymbol{\Gamma}[\hat{\rho}(\theta),\hat{\mathbf{\Pi}}_{\{r\}}]^{-1}\mathbf{d})^T\hat{\boldsymbol{\Pi}}_{\{r\}}+\alpha'\hat{\mathbb{I}}$ and the constants $\alpha,\alpha'\in\mathbb{R}$ have no influence on the sensitivity of the observable.

\end{document}